\documentclass{iauc}
\usepackage{graphicx}
\pubyear{2005}
\volume{199}
\pagerange{1--}
\jname{Probing Galaxies through Quasar Absorption Lines}
\editors{P. R. Williams, C. Shu, and B. M\'{e}nard, eds.}
\newcommand{\lya}{Lyman-$\alpha$~}

\title[The \lya forest as a probe of fundamental physics] 
{The \lya forest as a probe of fundamental physics}

\author[Matteo Viel]   
{Matteo Viel$^1$}

\affiliation{$^1$Institute of Astronomy, University of Cambridge,
Cambridge, CB3 0HA, UK \break 
email: viel@ast.cam.ac.uk}

\begin{document}

\maketitle

\begin{abstract}
We use LUQAS, a sample of 27 high resolution high signal-to-noise UVES
quasar (QSO) spectra (Kim et al. 2004), and the Croft et al. (2002)
sample together with a set of high resolution large box size
hydro-dynamical simulations run with the code GADGET-II to recover the
linear dark matter power spectrum at $z>2$ and at scales of $1-40 \,
h^{-1}\,$ Mpc. These scales cannot be probed by any other
observable. We address some of the uncertainties in the theoretical
modelling of the \lya forest structures such as the difference between
full hydro-dynamical simulations and a more simplified scheme based on
the HPM (Hydro-Particle-Mesh) technique.  We combine these data sets
with the Cosmic Microwave Background (CMB) data of WMAP in order to get
tighter constraints on cosmological parameters.  We focus on the
recovered values of the power spectrum amplitude, the primordial
spectral index and the running of the primordial spectral index. By
considering models of slow-roll inflation we give constraints on
inflationary parameters.  We explore implications for the mass of a
warm dark matter particle, including light gravitinos and sterile
neutrinos, and the perspectives of constraining the neutrino mass.

\end{abstract}

\section{Introduction}
The prominent absorption features blue-ward of the \lya emission in the
spectra of high-redshift quasars (QSOs) are now generally believed to
arise from smooth density fluctuations of a photoionised warm
intergalactic medium (e.g. \cite{rauch}).  This has opened up the
possibility to probe the density fluctuation of matter with the flux
power spectrum of QSO absorption lines (Croft et al. (1998); McDonald
et al. (2000); Gnedin \& Hamilton (2001); Hui et al. (2001); Croft et
al. (2002); McDonald et al. (2004); Viel, Haehnelt \& Springel
(2004)). The flux power spectrum is mainly sensitive to the slope and
amplitude of the linear dark matter power spectrum for wave-numbers in
the range $0.002\,<k$ (s/km) $< 0.05$, roughly corresponding to scales
1-40 comoving Mpc$/h$.  We use a suite of high-resolution
hydrodynamical simulations and the flux power spectrum obtained from
LUQAS (Large sample of UVES QSO Absorption Spectra), together with the
published flux power spectrum of Croft et al. (2002), to infer the
linear dark matter power spectrum at $z=2.125$ and $z=2.72$.  The idea
is to rely on accurate high resolution and large box-size
hydrodynamical simulations to model the bias function $b(k)$ which
relates the flux to the linear dark matter power spectrum: $P_F(k)=
b^2(k)\,P(k)$.  Moreover, we will briefly address some of the
uncertainties that arise from the theoretical (numerical) modelling of
the \lya forest structures.

We will then combine these estimates, in the framework of Monte Carlo
Markov Chains (\cite{cosmomc}), with the Cosmic Microwave Background
(CMB) results of WMAP (\cite{spergel}) to investigate the following
issues: {\it i)} recovery of cosmological parameters such as $\sigma_8$
(the power spectrum amplitude), $n_s$ and $n_{run}$ (the spectral index
and its running); {\it ii)} constraints on parameters describing
slow-roll inflationary models; {\it iii)} constraints on dark matter
particle masses and neutrinos.

\section{The LUQAS and the Croft et al. (2002) samples}
The LUQAS sample consists of 27 QSO spectra taken with the Ultra-Violet
Echelle Spectrograph (UVES) on VLT. Most of the spectra have been taken
as part of the Large ESO Observing programme UVESLP (P.I.:
J. Bergeron). The median redshift of the sample is $z=2.125$ and the
total redshift path is $\Delta z=13.75$. The typical signal-to-noise
ratio is $\sim 50$ and the pixel size is 0.05 \AA. For a more detailed
description of the sample and the data reduction we refer to
\cite{kim}. We combine this sample with that of Croft et al. (2002),
which consists of 30 Keck HIRES spectra and 23 Keck LRIS spectra and
has a median redshift of $z=2.72$. These two data sets seem to be in
agreement with the flux power spectrum obtained by
\cite{mcdonald2004,vielmoriond} from a large sample of low-resolution
SDSS spectra. Although some differences between the different data sets
are present, at the level of the error bars that we are going to quote,
these discrepancies are not relevant.  In all the data sets the
systematic uncertainties are larger than the statistical errors
(\cite{kim,mcdonald2004,vhs}).

\section{Hydrodynamical simulations: HPM vs full hydro}
\begin{figure}
\center\resizebox{.94\textwidth}{!}{\includegraphics{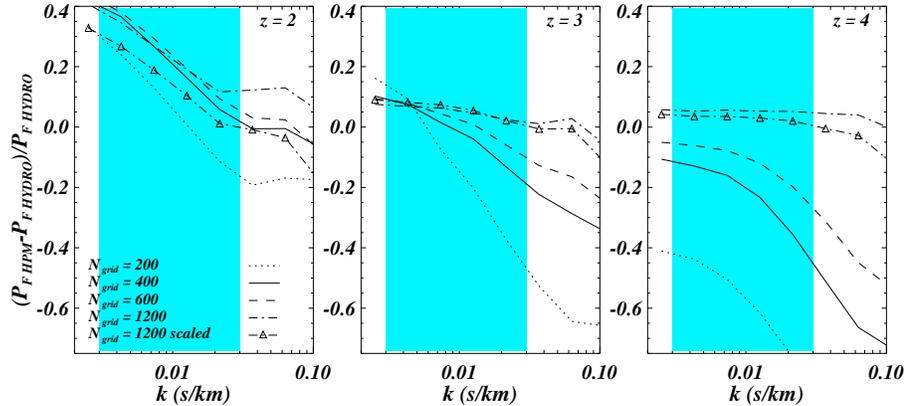}} 
  \caption{Fractional differences between the flux power spectra computed from
  full hydrodynamical simulations and HPM simulations with different
  resolutions of the grid size. The shaded area indicates the
  wavenumber range used for the recovery of the underlying linear dark
  matter power spectrum.}
\label{f1}
\end{figure}
Presently, one of the biggest source of error is due to the not very
well understood numerical modelling of the \lya forest (\cite{vhs}).
The Hydro-Particle-Mesh (HPM) method is an approximate technique to
simulate the low density IGM (\cite{gh}) and has been extensively used
to recover cosmological parameters from \lya forest data
(\cite{mcdonald2004,seljak}). This method is based on the numerical
implementation of an effective potential, that mimicks gas pressure,
into a particle mesh code that simulates collisionless dark matter.  We
decided to test the accuracy of this method in reproducing the flux
power spectrum, which is the quantity that is used to infer the dark
matter power spectrum, using state--of--the--art hydrodynamical
simulations run with the GADGET-II code.  The results are shown in
figure \ref{f1} where we plot the fractional differences between the
flux power spectrum of HPM simulations and the full hydrodynamical
one. We can see that if the resolution ($N_{\rm grid}$) of the HPM
method increases then the differences can be as small as 5-10\% at
$z=3,4$ (middle and left panels), in the wavenumber range of interest
in the analysis (represented by the shaded area).  At $z=2$ there are
strong discrepancies even at large scales that are due to the large
amount of shock-heated gas present in the full hydrodynamical
simulations that cannot be properly modelled with the HPM technique.
The line with overplotted empty triangles represents the case in which
the spectra have been scaled to match the same mean flux (both in the
HPM and full hydro runs), a procedure which is usually adopted by
simulators. We note that this scaling determines a better but still not
perfect agreement of the two flux power spectra.  Caution is thereby
needed when using HPM simulations to infer cosmological parameters and
an accurate calibration of these simulations with full hydrodynamical
ones is compelling (\cite{vhs2}).

\section{Results}
\subsection{Cosmological parameters}
We combined the linear dark matter power spectrum as inferred by
Viel, Haehnelt \& Springel (2004) with the CMB data of WMAP (\cite{spergel}). The results in
terms of cosmological parameters are here summarized: $\sigma_{8} =0.94
\pm 0.08$; $n_s=0.99 \pm 0.03$ (for models with no running, $1\sigma$
error bars); $n_s=0.959 \pm 0.036$ and $n_{run}=-0.033\pm 0.025$ (for a
model that includes running).  The data is thus consistent with a
scale-free power spectrum and no running of the spectral index.

For slow-roll inflationary models we get the following constraints:
$\epsilon_V = 0.032 \pm 0.018$, $\eta_V = 0.020 \pm 0.025$ and $\xi_V =
0.015 \pm 0.014$. The constraint on the tensor to scalar ratio is $r =
0.499 \pm 0.296$ (see \cite{vwh} for the relation between these
parameters and the spectral index and its running).  The results above
are in agreement with those of the SDSS collaboration (\cite{seljak}).

\subsection{Warm dark matter models}
Candidates of dark matter particles are generally classified according
to their velocity dispersion which defines a free-streaming length. On
scales smaller than the free-streaming length, fluctuations in the dark
matter density are erased and gravitational clustering is suppressed.
The velocity dispersion of Cold Dark Matter (CDM) particles is by
definition so small that the corresponding free-streaming length is
irrelevant for cosmological structure formation. That of Hot Dark
Matter, e.g. light neutrinos (see following section), is only one or
two orders of magnitude smaller than the speed of light, and smoothes
out fluctuations in the total matter density even on galaxy cluster
scales, which leads to strong bounds on their mass and density. Between
these two limits, there exists an intermediate range of dark matter
candidates generically called Warm Dark Matter (WDM). \cite{lesg} have
explored two different scenarios: pure warm dark matter models that
differ from $\Lambda$CDM models by a suppression of power below a scale
$\alpha$ (Figure 2); mixed models with CDM and gravitinos (or any warm
dark matter particles that decouples with a $g_* \sim 100$, shown in
Figure 3).

We obtain a lower limit (in the case of pure warm dark matter models)
of $m_{wdm} > 550$~eV ($2\sigma$) for early decoupled thermal relics
and $m_{wdm} > 2.0$ ~keV ($2\sigma$) for sterile neutrinos (that have
never been in thermal equilibrium).  In the case in which the gravitino
density is proportional to its mass we find an upper limit $m_{grav} <
16$~eV ($2\sigma$). This translates into a bound on the scale of
supersymmetry breaking, $\Lambda_{\rm susy} < 260$ TeV, for models of
supersymmetric gauge mediation in which the gravitino is the lightest
supersymmetric particle \cite{lesg}.

The values we get for the parameter $\alpha$ (which is related to the
scale of free streaming) and $f_x=\Omega_{grav}/\Omega_{DM}$ are the
following: $\alpha~(\mathrm{Mpc}/h) = 0.06 \pm 0.03$ and $f_x= 0.05\pm
0.04$ ($1\sigma$ error bars). Thereby, the concordance $\Lambda$CDM
model is perfectly consistent with the data.

\begin{figure}
\center\resizebox{0.58\textwidth}{!}{\includegraphics{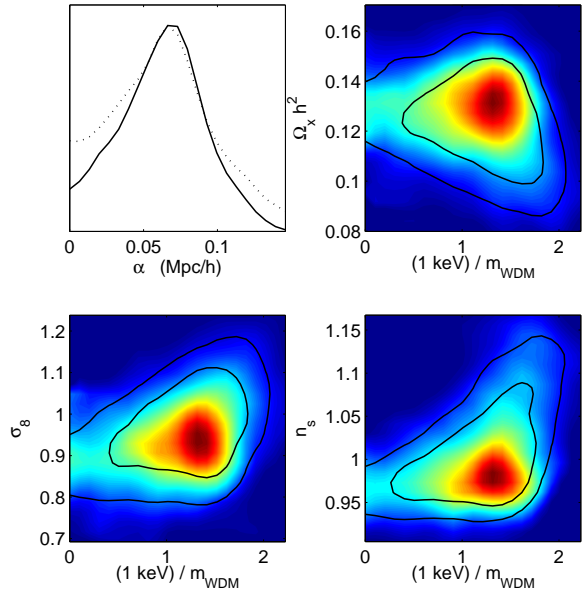}} 
  \caption{Scale of break for pure WDM thermal models. The power 
  is suppressed below the scale
  $\alpha$ (Mpc/h) and the suppression depends on the mass of the wdm
  particle. The line shows 1D-likelihood (upper-left panel, the
  marginalized result is shown as continous line); 2D-likelihood
  (1$\sigma$ and 2$\sigma$ contours) in
  colors (the lines represent the marginalized results while the
  coloured contours are for the mean likelihoods) for $\Omega_{WDM}$,
  $\sigma_8$ and $n_s$ vs the inverse of the wdm particle mass.
}
\label{f2}
\end{figure}
\begin{figure}
\center\resizebox{0.58\textwidth}{!}{\includegraphics{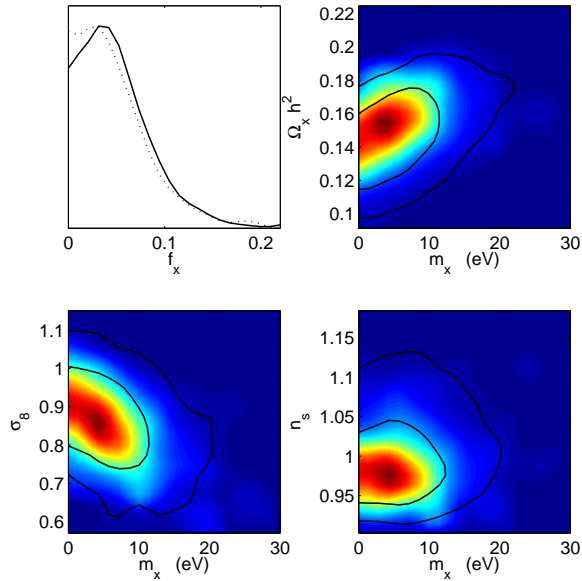}}
  \caption{Contribution of a light gravitino to the dark matter density
  $f_X=\Omega_{grav}/\Omega_{DM}$ (1D-likelihood, upper left panel). The
  plots are the same as in figure 2 but for a model of mixed cold+warm
  dark matter. Contour plots for the cosmological parameters are now
  plotted vs the gravitino mass (or any particles that decouples with a
  $g_*=100$).}
\label{f3}
\end{figure}

\subsection{Neutrino masses}
\begin{figure}
\center\resizebox{0.61\textwidth}{!}{\includegraphics{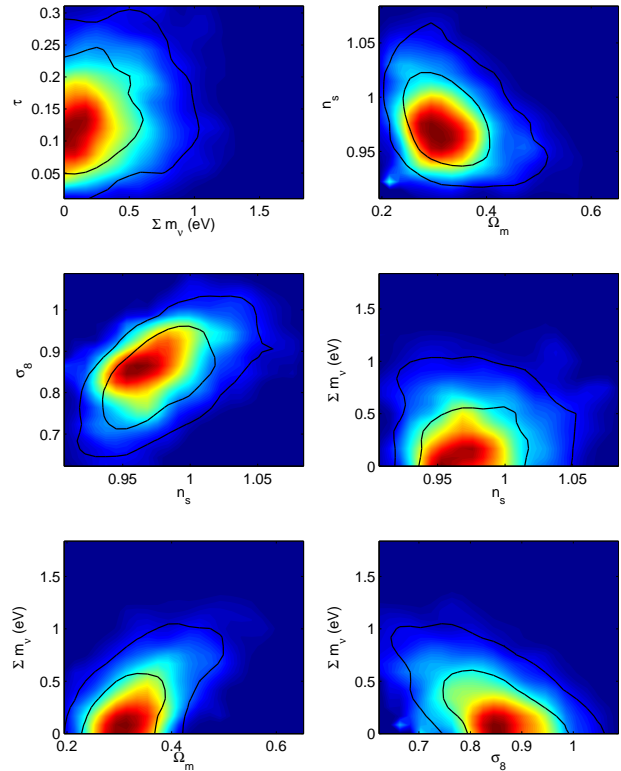}} 
  \caption{Constraints on the neutrino masses and on several
  cosmological parameters obtained by considering WMAP+2dF+\lya. Mean
  likelihoods and marginalized ones are represented by contours and
  lines, respectively.}
\label{f4}
\end{figure}

We have explored constraints on the quantity $\Sigma\,m_{\nu}(eV)$,
which is the sum of the mass of the light neutrino species. Both the
CMB and large-scale-structure are important tracers of neutrino
mass. At the time of decoupling neutrinos are still relativistic but
become non relativistic later in the evolution of the universe if their
mass is sufficiently high. As for WDM, but at larger scales, neutrinos
free-stream out of their potential wells, erasing perturbations on
smaller scales, producing a suppression of the amplitude, and leaving a
characteristic feature at the transition scale.  In Figure \ref{f4} we
plot 2D likelihood contours for several cosmological parameters. The
lines indicate the same likelihoods when marginalized over the full set
of parameters while the coloured contours are the mean likelihoods.
The results shown here have been obtained by using the WMAP data set,
the 2dF galaxy power spectrum (\cite{percival}) and the \lya
constraints of \cite{vhs}.  We obtain the following values for the
marginalized likelihoods: $\sigma_8=0.84\pm0.08$, $n_s=0.98\pm0.03$,
$\tau=0.15\pm0.06$, $\Omega_m=0.33\pm0.06$, $\Sigma m_{\nu}$ (eV)
=$0.33\pm0.27$ ($1\sigma$ error bars and for models which do not
consider the running of the spectral index and tensor contribution).

\section{Conclusions}\label{sec:concl}
The \lya forest, the absorption by neutral hydrogen seen in the spectra
of background QSOs, is a powerful laboratory for cosmology.  We have
used two samples of high resolution high signal-to-noise QSO spectra to
infer the linear dark matter power spectrum at $z>2$ and at scales not
probed by any other observables.  The present uncertainties in the
inferred values of the cosmological parameters are mainly due to the
systematic errors involved in the numerical modelling of the \lya
forest data with hydrodynamical simulations. We have explored the
differences between full hydrodynamical simulation and an approximate
scheme based on the HPM approximation.

We have then combined the small scales estimates of the \lya forest
with larger scales data sets such as the CMB and the 2dF galaxy power
spectrum.  The error bars on cosmological parameters that can be
inferred from the combined analysis of CMB and \lya forest data improve
by a factor two those obtained by the CMB data alone. The final results
point to a value of $\sigma_8=0.94\pm 0.08$ and a spectral index
$n_s=0.99\pm0.03$ ($1\sigma$ error bars) and no evidence for a running
of the spectral index. These results are consistent with those obtained
by the SDSS collaboration \cite{seljak}.  We have set constraints on
the mass of a warm (thermal) dark matter particle to be $m_{wdm} >550$
eV and on the light gravitino to be $m_{grav} < 16$ eV (2$\sigma$ error
bars). Further constraints on the sum of the three neutrinos can be
obtained and we get: $\Sigma m_{\nu}$ (eV) =$0.33\pm0.27$ ($1\sigma$
error bar) by combining WMAP, 2dF and \lya.

\begin{acknowledgments}
I thank my collaborators: Julien Lesgourgues, Martin Haehnelt, Sabino
Matarrese, Antonio Riotto, Volker Springel and Jochen Weller.  I thank
IAU for the award of a grant.  The simulations were run on the COSMOS
supercomputer at the Department of Applied Mathematics and Theoretical
Physics in Cambridge.
\end{acknowledgments}

\end{document}